\documentclass[a4paper,11pt]{article}
\pdfoutput=1 
\usepackage{jheppub} 
\usepackage[T1]{fontenc} 
\allowdisplaybreaks[1]	
\def\eqref#1{{Eq.\!~(\ref{#1})}} 
\def\figref#1{{Fig.\!~\ref{#1}}} 
\def\secref#1{{Sec.\!~\ref{#1}}} 
\def\ba#1\ea{\begin{align}#1\end{align}} 
\usepackage{scalerel}  
\usepackage{subfigure} 
\usepackage{bbold} 
\usepackage{mciteplus} 
\newcommand{\pperp}{\mathbin{\mathchoice
  {\xperp\scriptstyle}
  {\xperp\scriptstyle}
  {\xperp\scriptscriptstyle}
  {\xperp\scriptscriptstyle}
}}
\newcommand{\xperp}[1]{\vcenter{\hbox{$#1\perp$}}}
\usepackage{mathtools}

\title{\boldmath Topological charge fluctuations in the Glasma}

\author[a,b]{Pablo Guerrero-Rodríguez}

\affiliation[a]{CAFPE \& Dpto. de F\'{\i}sica Te\'orica y del Cosmos, \\ Universidad de Granada, E-18071 Campus de Fuentenueva, Granada, Spain.}
\affiliation[b]{Centre de Physique Th\'eorique, \'Ecole Polytechnique, \\
CNRS, Universit\'e Paris-Saclay, F-91128 Palaiseau, France.}

\emailAdd{pgr@ugr.es}

\abstract{The early-time evolution of the system generated in ultra-relativistic heavy ion collisions is dominated by the presence of strong color fields known as Glasma fields. These can be described following the classical approach embodied in the Color Glass Condensate effective theory, which approximates QCD in the high gluon density regime. In this framework we perform an analytical first-principles calculation of the two-point correlator of the divergence of the Chern-Simons current at proper time $\tau\!=\!0^+$, which characterizes the early fluctuations of axial charge density in the plane transverse to the collision axis. This object plays a crucial role in the description of anomalous transport phenomena such as the Chiral Magnetic Effect. We compare our results to those obtained under the Glasma Graph approximation, which assumes gluon field correlators to obey Gaussian statistics. While this approach proves to be equivalent to the exact calculation in the limit of short transverse separations, important differences arise at larger distances, where our expression displays a remarkably slower fall-off than the Glasma Graph result ($1/r^4$ vs.\ $1/r^8$ power-law decay). This discrepancy emerges from the non-linear dynamics mapping the Gaussianly-distributed color source densities onto the Glasma fields, encoded in the classical Yang-Mills equations. Our results support the conclusions reached in a previous work, where we found indications that the color screening of correlations in the transverse plane occurs at relatively large distances.}

\begin{document} 
\maketitle
\flushbottom

\section{Introduction}
\label{sec:intro}
High-energy heavy ion collisions (HICs from now on) generate a hot, chiral-symmetric medium --the Quark Gluon Plasma (QGP)-- where parity (P) and charge-parity (CP)-violating fluctuations are expected to happen with relatively high probability due to the chiral anomaly of QCD \cite{Fukushima:2008xe}. The anomalous term weights gauge field configurations according to their associated value of the topological invariant known as winding number (or topological charge):
\ba
Q_{\text{w}}=\frac{g^2}{16\pi^2}\int\!d^4x\,\text{Tr}\left\{F_{\mu\nu}(x)\tilde{F}^{\mu\nu}(x)\right\},
\ea
where $F^{\mu\nu}$ and $\tilde{F}^{\mu\nu}\!=\!\frac{1}{2}\epsilon^{\mu\nu\rho\sigma}F_{\rho\sigma}$ correspond respectively to the field strength tensor and its dual. $Q_{\text{w}}$ labels degenerate (but topologically inequivalent) vacuum states separated by potential barriers that greatly suppress the transition probability except at high temperatures, such as the ones reached in the QGP phase (for a review of the topological aspects of gauge field theories, see \cite{Lenz:2001me}). The tunneling transitions, mediated by localized field configurations called
 sphalerons, induce a transformation of left- into right-handed quarks (and viceversa) whose rate can be computed in the massless fermion limit as the spatial integral of the chiral anomaly:
\ba
\frac{dN_{5}}{dt}=\int\!d^3x\,\partial_{\mu}j^{\mu}_5(x)=-\frac{g^2N_f}{8\pi^2}\int\!d^3x\,\text{Tr}\left\{F_{\mu\nu}(x)\tilde{F}^{\mu\nu}(x)\right\}=\frac{g^2N_f}{2\pi^2}\int\!d^3x\,\dot{\nu}(x),\label{chiral}
\ea
where $N_{5}$ and $j_5^{\mu}$ correspond respectively to the axial charge density and current, $N_f$ is the number of flavors and $\dot{\nu}$ is the divergence of the Chern-Simons current.

HICs thus provide an appropriate environment --the QGP-- for the manifestation of the non-trivial topological structure of QCD. Specifically, off-central HICs yield favourable conditions for the search of observable signs of such features. These collisions give rise to large background electromagnetic fields, which in the presence of deconfined chirally-imbalanced matter may induce a separation of positive and negative charges along the direction of angular momentum \cite{Kharzeev:2007jp}. This effect --known as the Chiral Magnetic Effect (CME)-- thus creates a preferential direction for the emission of charged particles that would in turn translate into non-trivial azimuthal correlations in the hadronic spectrum. The search for such signatures of this and other transport phenomena connected to the chiral anomaly (generically called anomalous transport phenomena) has been carried at the Relativistic Heavy-Ion Collider (RHIC) and at the Large Hadron Collider (LHC) \cite{Kharzeev:2015znc}. Although these experiments have provided numerous measurements that are indeed compatible with said phenomena \cite{Abelev:2009ac,Abelev:2012pa,Adamczyk:2014mzf,Adamczyk:2015eqo,Tribedy:2017hwn}, the presence of large background effects (e.g.\ transverse momentum \cite{Bzdak:2010fd} and local charge \cite{Schlichting:2010qia} conservation, which also give rise to intrinsic back-to-back correlations; and final state interactions \cite{Ma:2011uma}) prevent from drawing definite conclusions. 
Hence, there is a strong interest from the high-energy QCD community in reducing this uncertainty. Significant advances have been achieved on the experimental side, including the development of different detection techniques \cite{Voloshin:2004vk,Bzdak:2012ia,Wen:2016zic,Xu:2017qfs} and, most recently, the implementation of an isobaric collision program at RHIC aimed at the isolation of CME backgrounds \cite{Huang:2018duz}. Still, a thorough approach to this task demands for better theoretical constraints on the dynamical origin of correlations between detected particles.

Non-trivial final state correlations in heavy ion experiments have been shown to reflect not only the collective dynamics of QGP, but also the event-by-event fluctuations that characterize the initial phase of the collisions \cite{Luzum:2013yya}. 
These early fluctuations
 provide a natural source of topological charge that competes with sphaleron transitions. Although the latter are known to dominate axial charge production in the QGP phase, throughout the pre-equilibrium stage both mechanisms are likely to yield a significant contribution\footnote{It has been argued that in the early stage of the collision sphaleron transitions are suppressed due to the boost invariance of the generated fields \cite{Kharzeev:2001ev}. However, as the system evolves towards thermalization and boost invariance wears down, they would be significantly enhanced \cite{Kharzeev:2001ev, Mace:2016svc}. Whether or not event-by-event fluctuations dominate over this or other mechanisms of axial charge production --like thermal fluctuations of the field strength-- is out of the scope of this paper. For more exhaustive discussions on this topic, the reader is referred to the aforementioned studies.}.
It is thus essential to understand and quantitatively constrain the influence of each source in the experimentally observed correlations.

However, there is currently no consensus about either the origin or the practical description of early event-by-event fluctuations. Different prescriptions arise from a wide variety of phenomenological models whose goal is to provide initial conditions for the quasi-ideal relativistic hydrodynamical evolution of QGP. The numerical values of the physical quantities required as input by these models are constrained by agreement with data, sometimes varying greatly from one model to another \cite{Luzum:2013yya, Retinskaya:2013gca}. Such discrepancy introduces a significant amount of uncertainty in both the precision and physical interpretation of most phenomenological studies of the expansion and cooling of QGP. This general issue is of particular importance for those hydrodynamical descriptions that mimic the effects induced by the chiral anomaly \cite{Newman:2005hd,Son:2009tf,Sadofyev:2010pr} since event-by-event fluctuations are expected to contribute a significant fraction of the initial axial charge densities. Therefore, a better theoretical control of the early stages of the collision is essential to properly characterize the origin and description of anomalous transport phenomena.

The Color Glass Condensate (CGC) effective theory (see e.g.\ \cite{Weigert:2005us,Gelis:2010nm} for a review) is arguably the most promising framework for the description of the early phase of HICs. CGC describes the high density of small-$x$ gluons carried by nuclei as strong color fields whose dynamics obey the classical Yang-Mills equations. The classical approximation is founded in the fact that for very large occupation numbers the quantum fluctuations represent a negligible correction to the strong background field.
This condition is at the base of the McLerran-Venugopalan model (MV model from now on) \cite{McLerran:1993ni,McLerran:1993ka,McLerran:1994vd}, where nuclei are represented by an ensemble of SU($N_c$) color charges that act as sources of the classical fields. The MV model thus assumes a separation of degrees of freedom that is performed at an arbitrary light-cone momentum $\Lambda^+$: particles with $p^+\!>\!\Lambda^+$ are taken as hard color charges (which represent the valence quarks), and these generate the small-$x$ dynamical modes (the strong color fields), which satisfy $p^+\!\!<\!\Lambda^+$. CGC incorporates the means to compute the quantum corrections to the MV model via the B-JIMWLK equations, which describe the evolution of the theory with $\Lambda^+$. This framework has been extensively applied in the description of the early, non-equilibrium stage of HICs known as Glasma phase \cite{Lappi:2006fp,Lappi:2006hq,Dumitru:2008wn,Fukushima:2011nq}.

In a previous work \cite{Albacete:2018bbv} we provided a first-principles calculation quantifying the size and extent of the transverse correlations of the energy-momentum tensor of Glasma at early times (and in a subsequent paper \cite{Giacalone:2019kgg} the results of said work were applied in the description of anisotropic flow harmonic coefficients to excellent agreement with data measured at both RHIC and LHC). In this follow-up paper we extend the classical treatment to the divergence of the Chern-Simons current:
\ba
\dot{\nu}(\tau=0^+,x_{\pperp})\equiv\dot{\nu}_{\scaleto{0}{4.2pt}}(x_{\pperp})=\text{Tr}\{E(\tau=0^+,x_{\pperp})B(\tau=0^+,x_{\pperp})\},\label{dotnudef}
\ea
where $E^{i}\!=\!-F^{0i}$ and $B^{k}\!=\!\frac{1}{2}\epsilon^{ijk}F^{ij}$ are, respectively, the Glasma chromo-electric and -magnetic fields, evaluated at an infinitesimal positive proper time $\tau\!=\!0^+$ at a point $x_{\pperp}$ of the plane transverse to the collision axis. This object is proportional to the strong CP-violating term of the QCD Lagrangian, which is the source for local production of axial charge (\eqref{chiral}). In the present work we evaluate the correlation function $\langle\dot{\nu}_{\scaleto{0}{4.2pt}}(x_{\pperp})\dot{\nu}_{\scaleto{0}{4.2pt}}(y_{\pperp})\rangle$, which characterizes the early event-by-event fluctuations of the $\dot{\nu}$ distribution. In evaluating this object we find relatively long-range correlations that contrast with both naive expectations one could have from the MV model --where we assume local correlations at the level of color source distributions-- and previously obtained results from the Glasma Graph approximation \cite{Lappi:2017skr} --which assumes a linear mapping of the statistics followed by the color sources onto the Glasma fields. Although such a discrepancy mirrors the results found in \cite{Albacete:2018bbv}, it is worth remarking that the calculations presented in this work yield an even larger difference with respect to those obtained under the Glasma Graph approximation.

This paper is organized as follows. In \secref{sec:setup} we introduce a generalization of the MV model with explicit impact parameter dependence and relaxed transversal locality. In this framework we outline the solution to the Yang-Mills equations with two sources at an infinitesimal proper time after the collision $\tau\!=\!0^+$, which acts as boundary condition for the ensuing evolution in the future light-cone. In \secref{sec:cs1} we calculate the expectation value of the divergence of the Chern-Simons current in the previously presented framework. In \secref{sec:cs2} we compute its two-point correlator, comparing our results with the aforementioned calculation performed under the Glasma Graph approximation \cite{Lappi:2017skr}. We also provide the first orders of the $N_c$-expansion, as well as the strict MV model limit of our expression. Note that a big part of the technical challenges faced during the calculations presented in this paper --such as the computation of non-trivial projections of the correlator of four Wilson lines in the adjoint representation and the decomposition of correlators of $m$ valence color sources and $n$ Wilson lines-- were analyzed in depth in a previous work \cite{Albacete:2018bbv}, being the reader referred to said paper for detailed derivations. Finally, in \secref{sec:end} we discuss the physical implications and potential phenomenological applications of our result, as well as its role in future studies.

\section{Setup: The classical approach to gluon production in high-energy heavy ion collisions}
\label{sec:setup}
In the MV model we represent the high density of small-$x$ gluons carried by each nucleus with gauge fields $A^{\mu}_{\scaleto{1,2}{5.8pt}}$ whose dynamics follow from the classical Yang-Mills equations:
\ba\label{YMeqs}
\left[ D_{\mu},F^{\mu\nu}\right]=\delta^{\nu+}\rho^{a}(x^{-},x_{\pperp})\,t^{a}\!,
\ea
where $\rho^a$ represents the density of valence (large-$x$) partons and $t^{a}$ is the generator of $SU(N_c)$ in the fundamental representation. The $\delta^{\nu+}$ factor indicates that the source generates a color current only in the + direction. This suggests a physical picture of the interaction where the valence partons do not recoil from their light-cone trajectory as the gluons they continuously exchange with the medium are too soft to affect their motion (eikonal approximation).

The MV model accounts for the event-by-event fluctuations of color charge configurations by taking $\rho^{a}$ as an stochastic quantity with a certain probability distribution $W[\rho]$ associated as weight function. Thus, the observables are obtained as ensemble averages over the background classical fields:
\ba\label{Average}
\langle {\cal O}[\rho] \rangle\!=\frac{1}{{\cal N}}\int[d\rho] W[\rho]{\cal O}[\rho],
\ea
where ${\cal N}$ is a normalization constant equal to $\int[d\rho] W[\rho]$. The main assumption adopted in the MV model is that in nuclei with large mass numbers the valence parton configurations emerge from a large number of separate nucleons and therefore are uncorrelated. Thus, invoking the central limit theorem, this model approximates $W[\rho]$ with a Gaussian distribution, yielding the following fundamental result:
\ba\label{MV2point}
\langle \rho^{a}(x^-,x_{\pperp})\rho^{b}(y^-,y_{\pperp})\rangle_{\scaleto{\text{MV}}{0.12cm}} = \mu^2(x^-)\delta^{ab}\delta(x^--y^-)\delta^2(x_{\pperp}-y_{\pperp}).
\ea
Here, $\mu^2$ is a parameter proportional to the color source number density that enters as the variance of the Gaussian weight. However, as we intend to apply a more general approach, we choose to relax some of the approximations implied in \eqref{MV2point} by considering the following, more general, two-point function of the sources:
\begin{align}
\langle \rho^{a}(x^-,x_{\pperp})\rho^{b}(y^-,y_{\pperp})\rangle &= \mu^2(x^-)h(b_{\pperp})\delta^{ab}\delta(x^--y^-)f(x_{\pperp}-y_{\pperp}) \nonumber\\
&\equiv\lambda(x^-,b_{\pperp})\delta^{ab}\delta(x^--y^-)f(x_{\pperp}-y_{\pperp}),\label{2pointMV}
\end{align}
where we allow the possibility of finite, non-homogeneous nuclei by explicitly introducing an impact parameter ($b_{\pperp}\!\equiv\!(x_{\pperp}\!+y_{\pperp})/2$) dependence as previously done in \cite{Chen:2015wia}. Also, we drop the assumption that interactions are local in the transversal plane by introducing an undetermined function $f(x_{\pperp}\!-y_{\pperp})$ instead of a Dirac delta. These extensions of the original MV model might prove especially useful in subsequent phenomenological applications of our results.

Although there is no general solution for the Yang-Mills equations with two sources, the MV model provides an analytical expression of the gauge fields at proper time $\tau\!=\!0^+$ (i.e.\ an infinitesimal positive proper time after the collision). These are obtained in terms of the gauge fields that characterize each nucleus before the collision, which can be computed independently as:
\begin{align}
A^{\pm}_{\scaleto{1}{4.2pt},\scaleto{2}{4.2pt}}(x^{\mp},x_{\pperp})&=0 \\
A^{i}_{\scaleto{1}{4.2pt},\scaleto{2}{4.2pt}}(x^{\mp},x_{\pperp})&=\theta(x^{\mp})\!\!\int^{\infty}_{-\infty}\!dz^{\mp}\int dz^2_{\pperp} G(z_{\pperp}\!-x_{\pperp})U_{\scaleto{1}{4.2pt},\scaleto{2}{4.2pt}}^{\dagger}(z^{\mp},x_{\pperp})\partial^{i}\tilde{\rho}_{\scaleto{1}{4pt},\scaleto{2}{4.2pt}}(z^{\mp},z_{\pperp})U_{\scaleto{1}{4.2pt},\scaleto{2}{4.2pt}}(z^{\mp},x_{\pperp})\nonumber\\
&\equiv\theta(x^{\mp})\alpha_{\scaleto{1}{4.2pt},\scaleto{2}{4.2pt}}^{i}(x_{\pperp}),
\end{align}
by solving the Yang-Mills equations with one source in the Fock-Schwinger gauge. Here $\tilde{\rho}$ is the color charge density in the covariant gauge and $U$ is the Wilson line, an SU($N_c$) element that represents the effect of the interaction with the classical gluon field over the fast valence partons in the eikonal approximation, i.e.\ a rotation in color space. It is defined as a path-ordered exponential:
\begin{align}
U_{\scaleto{1}{4.2pt},\scaleto{2}{4.2pt}}(x^{\mp},x_{\pperp})=\text{P}^{\mp} \exp{\left\{ -ig\!\int^{x^{\mp}}_{\scaleto{-\infty}{5pt}}\!\!dz^{\mp}\!\!\int dz^2_{\pperp}G(z_{\pperp}-x_{\pperp})\tilde{\rho}_{\scaleto{1}{4pt},\scaleto{2}{4pt}}(z^{\mp},z_{\pperp})\right\}},
\end{align}
where $G(z_{\pperp}\!\!-\!x_{\pperp})$ is the Green's function for the two-dimensional Laplace operator. The solution at $\tau\!=\!0^+$ is found by proposing the following ansatz:
\begin{align}
A^{\pm}(x^{\mp},x_{\pperp})&\!=\!\pm \,x^{\pm} \alpha(\tau=0^+\!,x_{\pperp})\label{Ansatz1}\\
A^{i}\;(x^{\mp},x_{\pperp})&\!=\!\alpha^{i}(\tau=0^+\!,x_{\pperp}),\label{Ansatz}
\end{align}
where we adopted the comoving coordinate system, defined by proper time $\tau\!=\!\sqrt{2x^+x^-}$ and rapidity $\eta\!=\!\frac{1}{2}\ln (x^+/x^-)$. Then, we invoke a physical `matching condition' that requires Yang-Mills equations to be regular in the limit $\tau\!\rightarrow\!0$
. In doing so, the following relations are obtained:
\begin{align}
\alpha^{i} (\tau=0^+\!,x_{\pperp})&= \alpha_{\scaleto{1}{4.2pt}}^{i}(x_{\pperp})+\alpha_{\scaleto{2}{4.2pt}}^{i}(x_{\pperp})\label{EvolutionBoundaries1}\\
\alpha (\tau=0^+\!,x_{\pperp})&= \frac{ig}{2} \left[ \alpha_{\scaleto{1}{4.2pt}}^{i}(x_{\pperp}) , \alpha_{\scaleto{2}{4.2pt}}^{i}(x_{\pperp}) \right]\!,\label{EvolutionBoundaries}
\end{align}
which act as boundary conditions of the subsequent $\tau$-evolution\footnote{Several approaches of both analytical and numerical nature have been applied for this computation in the literature. For instance, in \cite{Fries:2006pv} an analytical approximation based on an expansion of the Yang-Mills equations in powers of $\tau$ is proposed.
However, this is out of the scope of the work presented in this paper.}. From these expressions we can compute the chromo-electric and -magnetic fields at $\tau\!=\!0^+$ as:
\ba
E^z(\tau=0^+,x_{\pperp})=&-ig\,\delta^{ij}[\alpha^{i}_{\scaleto{1}{4.2pt}}(x_{\pperp}),\alpha^{j}_{\scaleto{2}{4.2pt}}(x_{\pperp})]\label{elmag1}\\
B^z(\tau=0^+,x_{\pperp})=&-ig\,\epsilon^{ij}[\alpha^{i}_{\scaleto{1}{4.2pt}}(x_{\pperp}),\alpha^{j}_{\scaleto{2}{4.2pt}}(x_{\pperp})],\label{elmag2}
\ea
being these the only non-vanishing components. This peculiar configuration of boost-invariant longitudinal fields motivates the Glasma flux tube picture, which predicts short-range transverse correlations \cite{Iancu:2002aq}.

\section{One-point correlator of the divergence of the Chern-Simons current} 
\label{sec:cs1}
Before evaluating the two-point function, in this section we will show that the expectation value of the divergence of the Chern-Simons current over the classical background fields is 0, indicating that there is no overall CP violation in the process.
Although this result has been obtained previously in the literature \cite{Lappi:2006fp,Lappi:2017skr}, we deem it convenient to include this preface as it allows us to establish the notation used in the rest of the paper. Let us first write $\dot{\nu}_{\scaleto{0}{4.2pt}}$ in terms of the gluon fields:
\ba
\dot{\nu}_{\scaleto{0}{4.2pt}}(x_{\pperp})&=-g^2\delta^{ij}\epsilon^{kl}\text{Tr}\{[\alpha^{i}_{\scaleto{1}{4.2pt}\,x},\alpha^{j}_{\scaleto{2}{4.2pt}\,x}][\alpha^{k}_{\scaleto{1}{4.2pt}\,x},\alpha^{l}_{\scaleto{2}{4.2pt}\,x}]\}\nonumber\\
&=-g^2\delta^{ij}\epsilon^{kl}\alpha^{i,a}_{\scaleto{1}{4.2pt}\,x}\alpha^{j,b}_{\scaleto{2}{4.2pt}\,x}\alpha^{k,c}_{\scaleto{1}{4.2pt}\,x}\alpha^{l,d}_{\scaleto{2}{4.2pt}\,x}\text{Tr}\{[t^{a},t^{b}][t^{c},t^{d}]\}\nonumber\\
&=\frac{g^2}{2}\delta^{ij}\epsilon^{kl}f^{abn}f^{cdn}\alpha^{i,a}_{\scaleto{1}{4.2pt}\,x}\alpha^{j,b}_{\scaleto{2}{4.2pt}\,x}\alpha^{k,c}_{\scaleto{1}{4.2pt}\,x}\alpha^{l,d}_{\scaleto{2}{4.2pt}\,x},\label{corree}
\ea
where we adopted the shorthand notation $\alpha^{i,a}(x_{\pperp})\!\equiv\!\alpha^{i,a}_{x}$. As an intermediate step we expand the color structure of the gluon fields as:
\ba
\alpha^{i}(x_{\pperp})\!&=\!\int^{\infty}_{-\infty}\!dz^-\!\!\int dz^2_{\pperp}\,G(z_{\pperp}\!-x_{\pperp})\partial^{i}\tilde{\rho}^{a}(z^-,z_{\pperp})U^{\dagger}(z^-,x_{\pperp})t^{a}U(z^-,x_{\pperp})\!\nonumber\\
&=\!\int^{\infty}_{-\infty}dz^-\!\!\int dz^2_{\pperp}\,G(z_{\pperp}\!-x_{\pperp})\partial^{i}\tilde{\rho}^{a}(z^-,z_{\pperp})U^{ab}(z^-,x_{\pperp})t^{b}\equiv \alpha^{i,b}(x_{\pperp}) t^{b},
\ea
where we used the relation between Wilson lines in the fundamental and adjoint representations $U^{\dagger}t^{a}U\!=\!U^{ab}t^{b}$.
The correlator of \eqref{corree} factorizes like:
\ba
\langle\dot{\nu}_{\scaleto{0}{4.2pt}}(x_{\pperp})\rangle=\frac{g^2}{2}\delta^{ij}\epsilon^{kl}f^{abn}f^{cdn}\langle\alpha^{i,a}_{\scaleto{1}{4.2pt}}(x_{\pperp})\alpha^{k,c}_{\scaleto{1}{4.2pt}}(x_{\pperp})\rangle\langle\alpha^{j,b}_{\scaleto{2}{4.2pt}}(x_{\pperp})\alpha^{l,d}_{\scaleto{2}{4.2pt}}(x_{\pperp})\rangle,\label{Correlator1}
\ea
as in the MV model the color source fluctuations of each nucleus are assumed to be independent. 
Thus, the building block of $\langle\dot{\nu}_{\scaleto{0}{4.2pt}}\rangle$ is the correlator of two gauge fields evaluated in the same transverse position, $\langle\alpha^{i,a}(x_{\pperp})\alpha^{k,c}(x_{\pperp})\rangle$. We calculate this object as a limit:
\ba
\left\langle \alpha^{i,a}(x_{\pperp})\alpha^{j,b}(x_{\pperp}) \right\rangle&=\delta^{ab}\lim_{r\rightarrow0}\!\int^{\infty}_{-\infty}dz^-\lambda(z^-,b_{\pperp})\partial^{i}_{x}\partial^{j}_{y}L(r_{\pperp})C^{(2)}_{\text{adj}}(z^-;x_{\pperp},y_{\pperp}),\label{Resultxy}
\ea
where $r\!=\!|r_{\pperp}|\!=\!|x_{\pperp}-y_{\pperp}|$. Here we introduced the following function:
\ba
L(r_{\pperp})\equiv\int\! dz^2_{\pperp}du^2_{\pperp}G(z_{\pperp}\!-x_{\pperp})G(u_{\pperp}\!-y_{\pperp})f(z_{\pperp}\!-u_{\pperp}).
\ea
From its symmetries and dimension, the double derivative of $L(r_{\pperp})$ featured in \eqref{Resultxy} can be parameterized as:
\ba
\partial^{i}_{x}\partial^{j}_{y}L(r_{\pperp})=A(r_{\pperp})\delta^{ij}+B(r_{\pperp})\!\left( \frac{\delta^{ij}}{2}-\frac{r^{i}r^{j}}{r^2}\right)\!.\label{param}
\ea
This formula accounts for an explicit separation of the contributions of the unpolarized $A(r_{\pperp})$ and linearly polarized $B(r_{\pperp})$ parts of the gluon distribution. We can express these coefficients in terms of $f(r_{\pperp})$ explicitly as:
\ba
A(r_{\pperp})&=\!\frac{1}{2}\!\int\!\frac{d^2q_{\pperp}}{(2\pi)^2}\hat{f}(q_{\pperp})\frac{e^{iq_{\pperp}\!\cdot r_{\pperp}}}{q^2}\label{unpol}\\
B(r_{\pperp})&=-\int\!\frac{d^2q_{\pperp}}{(2\pi)^2}\hat{f}(q_{\pperp})\frac{e^{iq\,r\cos\theta}}{q^2}\cos(2\theta),\label{pol}
\ea
where $\hat{f}(q_{\pperp})$ is the Fourier transform of $f(r_{\pperp})$.
As for the last factor in \eqref{Resultxy}, it corresponds to the dipole function in the adjoint representation, which stems from the following correlator:
\ba
\left\langle U^{a^{\prime}a}(x^-,x_{\pperp})U^{a^{\prime}b}(x^-,y_{\pperp}) \right\rangle\!&=\delta^{ab}\exp{\left\{ -g^2\frac{N_c}{2}\Gamma(r_{\pperp})\bar{\lambda}(x^-,b_{\pperp})\right\}}\nonumber\\
&\equiv\delta^{ab}C^{(2)}_{\text{adj}}(x^-;x_{\pperp},y_{\pperp}).\label{dipolefun}
\ea
Here we introduced the notation $\bar{\lambda}(x^-,x_{\pperp})\!=\!\int^{x^-}_{-\infty}dz^-\lambda(z^-,x_{\pperp})$, as well as the following function:
\ba
\Gamma(r_{\pperp})\!=\!2(L(0_{\pperp})-L(r_{\pperp})).
\ea
Computing the indicated limit explicitly, \eqref{Resultxy} yields:
\ba
\left\langle \alpha^{i,a}(x_{\pperp})\alpha^{j,b}(x_{\pperp}) \right\rangle=-\frac{1}{2}\delta^{ab}\delta^{ij}\bar{\mu}^2h(x_{\pperp})\partial^2L(0_{\pperp})=-\frac{1}{2}\delta^{ab}\delta^{ij}\bar{\lambda}(x_{\pperp})\partial^2L(0_{\pperp}).\label{Result11}
\ea
(For a more detailed calculation of this correlator we refer the reader to \cite{Albacete:2018bbv}). The factor $\bar{\lambda}(x_{\pperp})$ corresponds to $\lambda$ integrated from $-\infty$ to $\infty$ in the longitudinal direction (in general, we will identify functions integrated in the longitudinal direction from $-\infty$ to $\infty$ by simply omitting their longitudinal dependence). The factor $\partial^2L(0_{\pperp})$ is a model-dependent constant that emerges from the following limit:
\ba
\lim_{r\rightarrow0}\partial^{i}_{x}\partial^{j}_{y}L(r_{\pperp})=\frac{\delta^{ij}}{2}\int\frac{d^2q_{\pperp}}{(2\pi)^2}\hat{f}(q_{\pperp})\frac{1}{q^2}\equiv-\frac{1}{2}\delta^{ij}\partial^2L(0_{\pperp}).
\ea
Some assumptions had to be made about the functions $h(b_{\pperp})$, $f(r_{\pperp})$ introduced in \eqref{2pointMV} in order to arrive at the expressions presented in this section. Specifically, we take $h(b_{\pperp})$ as a slowly varying function over lengths of the order of an infrared length scale $1/m$ (or smaller), which can be understood as a cut-off that imposes color neutrality at the nucleon size. In addition, we assume that $f(r_{\pperp})$ behaves in such a way that its Fourier transform $\hat{f}(k_{\pperp})$ tends to unity in the infrared limit (see \cite{Albacete:2018bbv} for details).

When finally substituting our building block \eqref{Result11} into \eqref{Correlator1}, we get:
\ba
\langle\dot{\nu}_{\scaleto{0}{4.2pt}}(x_{\pperp})\rangle=\frac{g^2}{8}(\partial^2L(0_{\pperp}))^2\bar{\lambda}_{\scaleto{1}{4.2pt}}(x_{\pperp})\bar{\lambda}_{\scaleto{2}{4.2pt}}(x_{\pperp})\delta^{ij}\epsilon^{kl}f^{abn}f^{cdn}\delta^{ac}\delta^{ik}\delta^{bd}\delta^{jl}=0,
\ea
which vanishes due to the antisymmetric property of the Levi-Civita tensor.
As mentioned earlier, this null average accounts for the Glasma state being globally CP-symmetric. However, as will be shown below, local axial charge fluctuations are expected to happen on an event-by-event basis. Our object of interest is therefore the two-point correlator of $\dot{\nu}_{\scaleto{0}{4.2pt}}$, whose computation we outline in the following section.

\section{Two-point correlator of the divergence of the Chern-Simons current} 
\label{sec:cs2}
\begin{figure}
\centering
\includegraphics[width=0.528\textwidth]{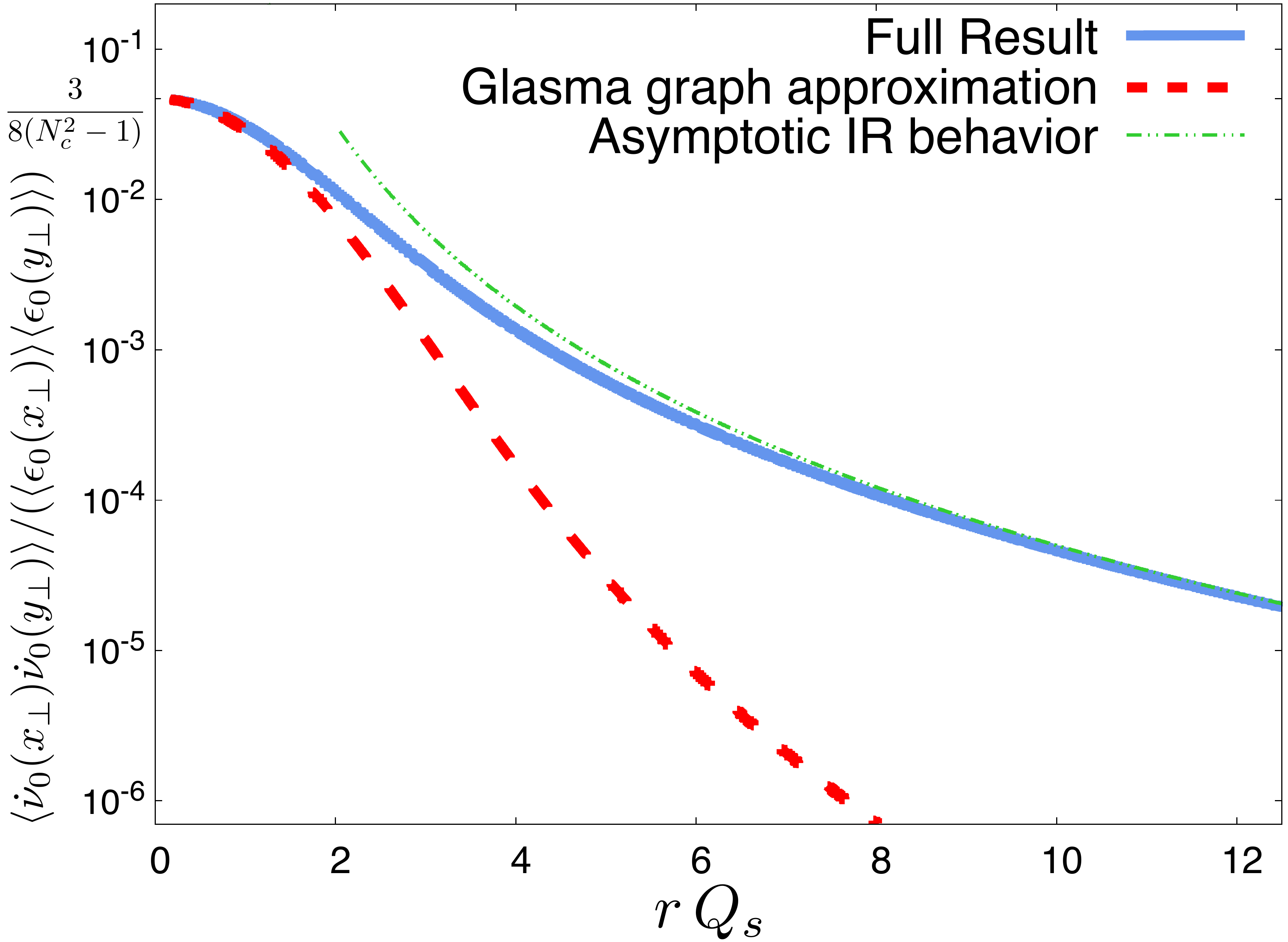}
\caption{Two-point correlator of the divergence of the Chern-Simons current normalized to the product of energy density averages in the exact analytical approach (blue full curve) and the Glasma Graph approximation (red dashed curve). As a visual aid we also indicate the asymptotic behavior in the infrared limit, which is $32/[(N_c^2-1)^2r^4Q_s^4]$ (green dot-dashed curve).}
\label{Result1}
\end{figure}

In this section we describe the calculation of $\langle\dot{\nu}_{\scaleto{0}{4.2pt}}(x_{\pperp})\dot{\nu}_{\scaleto{0}{4.2pt}}(y_{\pperp})\rangle$, which characterizes the early fluctuations of the divergence of the Chern-Simons current in the transverse plane. As we did in the previous section, we start by expanding $\dot{\nu}_{\scaleto{0}{4.2pt}}$ in terms of the gluon fields:
\ba
\dot{\nu}_{\scaleto{0}{4.2pt}}(x_{\pperp})\dot{\nu}_{\scaleto{0}{4.2pt}}(y_{\pperp})=\frac{g^4}{4}\delta^{ij}\epsilon^{kl}\delta^{i'j'}\!\epsilon^{k'l'}\!f^{abn}f^{cdn}f^{a'b'm}f^{c'd'm}\alpha^{i,a}_{1x}\alpha^{j,b}_{2x}\alpha^{k,c}_{1x}\alpha^{l,d}_{2x}\alpha^{i',a'}_{1y}\!\alpha^{j',b'}_{2y}\!\alpha^{k',c'}_{1y}\!\alpha^{l',d'}_{2y},
\ea
then, the correlator reads:
\ba
\langle\dot{\nu}_{\scaleto{0}{4.2pt}}(x_{\pperp})\dot{\nu}_{\scaleto{0}{4.2pt}}(y_{\pperp})\rangle=\frac{g^4}{4}\epsilon^{kl}\epsilon^{k'l'}f^{abn}f^{cdn}f^{a'b'm}f^{c'd'm}\langle\alpha^{i,a}_x\alpha^{k,c}_x\alpha^{i',a'}_y\!\alpha^{k',c'}_y\rangle_{\scaleto{1}{4.2pt}}\langle\alpha^{i,b}_x\alpha^{l,d}_x\alpha^{i',b'}_y\!\alpha^{l',d'}_y\rangle_{\scaleto{2}{4.2pt}}.
\ea
Color algebra-wise, this expression presents the same level of complexity than the two-point correlator of the energy density, previously computed in \cite{Albacete:2018bbv}. Happily, it features a much simpler transverse index structure, which we can rewrite as:
\ba
\epsilon^{kl}\epsilon^{k'l'}=\delta^{kk'}\delta^{ll'}\!-\delta^{kl'}\delta^{lk'},
\ea
yielding:
\ba
\langle\dot{\nu}_{\scaleto{0}{4.2pt}}(x_{\pperp})\dot{\nu}_{\scaleto{0}{4.2pt}}(y_{\pperp})\rangle=\frac{g^4}{4}f^{abn}f^{cdn}f^{a'b'm}f^{c'd'm}\Big(&\langle\alpha^{i,a}_x\alpha^{k,c}_x\alpha^{i',a'}_y\!\alpha^{k,c'}_y\rangle_{\scaleto{1}{4.2pt}}\langle\alpha^{i,b}_x\alpha^{l,d}_x\alpha^{i',b'}_y\!\alpha^{l,d'}_y\rangle_{\scaleto{2}{4.2pt}}\nonumber\\
&-\langle\alpha^{i,a}_x\alpha^{k,c}_x\alpha^{i',a'}_y\!\alpha^{l,c'}_y\rangle_{\scaleto{1}{4.2pt}}\langle\alpha^{i,b}_x\alpha^{l,d}_x\alpha^{i',b'}_y\!\alpha^{k,d'}_y\rangle_{\scaleto{2}{4.2pt}}\Big).\label{2point}
\ea
The building block for this computation is the correlator of four gluon fields in two different transverse positions:
\ba
\langle \alpha^{i,a}(x_{\pperp}) \alpha^{k,c}(x_{\pperp}) \alpha^{i^{\prime}\!,a^{\prime}}(y_{\pperp})\alpha^{k^{\prime}\!,c^{\prime}}(y_{\pperp}) \rangle\!=\!\!\int^{\infty}_{-\infty}\!dz^-dw^-dz^{-\prime}dw^{-\prime}
\bigg\langle \!\frac{\partial^{i}\tilde{\rho}^{e}(z^-,x_{\pperp})}{\nabla^2}U^{ea}(z^-,x_{\pperp})\nonumber\\
\frac{\partial^{k}\tilde{\rho}^{f}(w^-,x_{\pperp})}{\nabla^2}U^{fc}(w^-,x_{\pperp})\frac{\partial^{i^{\prime}}\tilde{\rho}^{e^{\prime}}\!(z^{-\prime},y_{\pperp})}{\nabla^2}U^{e'a'}\!(z^{-\prime},y_{\pperp})\frac{\partial^{k^{\prime}}\tilde{\rho}^{f^{\prime}}\!(w^{-\prime},y_{\pperp})}{\nabla^2}U^{f'\!c'}\!(w^{-\prime},y_{\pperp}) \!\bigg\rangle,
\ea
where $1/\nabla^2$ is the shorthand notation we adopt to denote a convolution with the Laplacian Green's function. The above correlator is a highly non-trivial object whose calculation poses a number of outstanding technical challenges such as the computation of non-trivial projections of the correlator of four Wilson lines in the adjoint representation and the decomposition of correlators of four valence color sources and four Wilson lines. 
For a detailed calculation the reader is referred to \cite{Albacete:2018bbv}. Here we will simply indicate the result, for which we need to make some definitions. We start with the `connected' function:
\ba
C^{ij;kl}_{ab;cd}(x_{\pperp},&\,y_{\pperp},x_{\pperp},y_{\pperp})\!=f^{ace}f^{bde}\partial^{i}_{x}\partial^{j}_{y}L(x_{\pperp}\!-y_{\pperp})\partial^{k}_{x}\Gamma(x_{\pperp}-y_{\pperp})\partial^{l}_{y}\Gamma(y_{\pperp}-x_{\pperp})\nonumber\\
&\times\!\left( \frac{4}{\Gamma ^3 g^4 N_c^3}-\left(\frac{\bar{\lambda}^2(b_{\pperp}) }{2\Gamma N_c}+\frac{4}{\Gamma ^3 g^4 N_c^3}+\frac{2 \bar{\lambda}(b_{\pperp})}{\Gamma ^2 g^2 N_c^2}\right)\!C_{\text{adj}}^{(2)}(x_{\pperp},y_{\pperp})\right)\!,\label{connectedf}
\ea
which accounts for the contribution of correlations between the `external' color source densities and those arranged inside the Wilson lines. The remaining terms stem from a complete factorization of the correlations of color source densities and Wilson lines:
 \ba
\int^{\infty}_{-\infty}dz^-dw^-dz^{-\prime}dw^{-\prime}&\left\langle \frac{\partial^{i}\tilde{\rho}^{e}(z^-,x_{\pperp})}{\nabla^2}\frac{\partial^{k}\tilde{\rho}^{f}(w^-,x_{\pperp})}{\nabla^2}\frac{\partial^{i^{\prime}}\!\tilde{\rho}^{e^{\prime}}\!(z^{-\prime},y_{\pperp})}{\nabla^2}\frac{\partial^{k^{\prime}}\!\tilde{\rho}^{f^{\prime}}\!(w^{-\prime},y_{\pperp})}{\nabla^2}\right\rangle\nonumber\\
&\times\!\left\langle U^{ea}(z^-,x_{\pperp})U^{fc}(w^-,x_{\pperp})U^{e'a'}\!(z^{-\prime},y_{\pperp})U^{f'\!c'}\!(w^{-\prime},y_{\pperp}) \right\rangle.\label{Disconnected1}
 \ea
These contributions are described in terms of the two `disconnected' functions:
\ba
D^{ik;i'k'}_{ac;a'c'}(x_{\pperp},x_{\pperp},y_{\pperp},y_{\pperp})\!=\frac{1}{4}\delta^{ik}\delta^{i'k'}\!\!\left(\partial^2L(0_{\pperp})\right)^2\!\delta^{ac}\delta^{a^{\prime}c^{\prime}}\bar{\lambda}^2(b_{\pperp}),
\ea
and:
\ba
\hspace{-1cm}D^{ij;kl}_{ab;cd}(x_{\pperp},y_{\pperp},x_{\pperp},y_{\pperp})\!=\!2\!\left(\!\delta^{ab}\delta^{cd}\!\left[\frac{N_c^2-4}{2N_c^2}f_{\scaleto{1}{4pt}}+\frac{2}{N_c^2}f_{\scaleto{2}{4pt}}+\frac{N_c+2}{4N_c}f_{\scaleto{3}{4pt}}+\frac{N_c-2}{4N_c}f_{\scaleto{4}{4pt}}\right]\right.\nonumber\\
+\delta^{ac}\delta^{bd}\!\left[\frac{1}{N_c^2-1}f_{\scaleto{5}{4pt}}-\frac{N_c+2}{2N_c(N_c+1)}f_{\scaleto{3}{4pt}}+\frac{N_c-2}{2N_c(N_c-1)}f_{\scaleto{4}{4pt}}\right]\nonumber\\
+\delta^{ad}\delta^{bc}\!\left[-\frac{N_c^2-4}{2N_c^2}f_{\scaleto{1}{4pt}}-\frac{2}{N_c^2}f_{\scaleto{2}{4pt}}+\frac{N_c+2}{4N_c}f_{\scaleto{3}{4pt}}+\frac{N_c-2}{4N_c}f_{\scaleto{4}{4pt}}\right]\nonumber\\
+d^{abm}d^{cdm}\!\left[-\frac{1}{N_c}f_{\scaleto{1}{4pt}}+\frac{1}{N_c}f_{\scaleto{2}{4pt}}+\frac{1}{4}f_{\scaleto{3}{4pt}}-\frac{1}{4}f_{\scaleto{4}{4pt}}\right]+d^{adm}d^{cbm}\!\left[ \frac{1}{N_c}f_{\scaleto{1}{4pt}}-\frac{1}{N_c}f_{\scaleto{2}{4pt}}+\frac{1}{4}f_{\scaleto{3}{4pt}}-\frac{1}{4}f_{\scaleto{4}{4pt}}\right]\nonumber\\
\left.+d^{acm}d^{bdm}\!\left[\frac{N_c}{N_c^2-4}f_{\scaleto{2}{4pt}}-\frac{N_c+4}{4(N_c+2)}f_{\scaleto{3}{4pt}}+\frac{N_c-4}{4(N_c-2)}f_{\scaleto{4}{4pt}}\right]\,\right)\!T^{\,ij;kl}\!(x_{\pperp},y_{\pperp};x_{\pperp},y_{\pperp}),\label{DisconResult}
\ea
where:
\ba
    f_{\scaleto{1}{4pt}}=&\frac{2}{(N_cg^2\Gamma)^2}(1-C^{(2)}_{\text{adj}}(x_{\pperp},y_{\pperp}))^2\\
    f_{\scaleto{2}{4pt}} =&\frac{2}{N_cg^2\Gamma}\left( \frac{2}{N_cg^2\Gamma}(1-C^{(2)}_{\text{adj}}(x_{\pperp},y_{\pperp}))-\bar{\lambda}(b_{\pperp})C^{(2)}_{\text{adj}}(x_{\pperp},y_{\pperp})\right)\\
    f_{\scaleto{3}{4pt}} =& \left( \frac{4}{N_c(N_c+2)g^4\Gamma^2}(1-C^{(2)}_{\text{adj}}(x_{\pperp},y_{\pperp}))\right.\nonumber\\
   &\left.-\frac{2}{(N_c+2)(N_c+1)g^4\Gamma^2}\Big(1-(C^{(2)}_{\text{adj}}(x_{\pperp},y_{\pperp}))^2\exp{\left\{-g^2\Gamma \bar{\lambda}(b_{\pperp}) \right\}}\Big)\right)\\
    f_{\scaleto{4}{4pt}} =& \left( \frac{4}{N_c(N_c-2)g^4\Gamma^2}(1-C^{(2)}_{\text{adj}}(x_{\pperp},y_{\pperp}))\right.\nonumber\\
    &\left.-\frac{2}{(N_c-2)(N_c-1)g^4\Gamma^2}\Big(1-(C^{(2)}_{\text{adj}}(x_{\pperp},y_{\pperp}))^2\exp{\left\{g^2\Gamma \bar{\lambda}(b_{\pperp}) \right\}}\Big)\right)\\
    f_{\scaleto{5}{4pt}} =& \frac{2}{N_cg^2\Gamma}\left( \bar{\lambda}(b_{\pperp})-\frac{2}{N_cg^2\Gamma}(1-C^{(2)}_{\text{adj}}(x_{\pperp},y_{\pperp}))\right)\!.
\ea
The remarkable length of this function is a consequence of an specific step of its derivation process, namely the propagation of a non-trivial color state via a $6\!\times\!6$ matrix that represents the correlator of four Wilson lines in the adjoint representation in color space (see \cite{Albacete:2018bbv} for details).
Having defined these functions, we can write our building block compactly as:
\ba
\langle \alpha^{i\,a}(x_{\pperp}) \alpha^{k\,c}(x_{\pperp}) \alpha^{i^{\prime}a^{\prime}}\!(y_{\pperp})\alpha^{k^{\prime}c^{\prime}}\!(y_{\pperp}) \rangle= D^{ik;i'k'}_{ac;a'c'}(x_{\pperp},x_{\pperp},y_{\pperp},y_{\pperp})+D^{ii';kk'}_{aa';cc'}(x_{\pperp},y_{\pperp},x_{\pperp},y_{\pperp})\nonumber\\
+D^{ik';ki'}_{ac';ca'}(x_{\pperp},y_{\pperp},x_{\pperp},y_{\pperp})+C^{ii';kk'}_{aa';cc'}(x_{\pperp},y_{\pperp},x_{\pperp},y_{\pperp})+C^{ik';ki'}_{ac';ca'}(x_{\pperp},y_{\pperp},x_{\pperp},y_{\pperp})\nonumber\\
+C^{kk';ii'}_{cc';aa'}(x_{\pperp},y_{\pperp},x_{\pperp},y_{\pperp})+C^{ki';ik'}_{ca';ac'}(x_{\pperp},y_{\pperp},x_{\pperp},y_{\pperp}).\label{MainCorrelator2}
\ea
Substituting this result in \eqref{2point} and performing the ensuing index contractions (for which we use the Mathematica package FeynCalc \cite{Mertig:1990an,Shtabovenko:2016sxi}), we obtain the main result of this work:
\vspace{-0.07cm}
\ba
\langle&\dot{\nu}_{\scaleto{0}{4.2pt}}(x_{\pperp})\dot{\nu}_{\scaleto{0}{4.2pt}}(y_{\pperp})\rangle=\frac{16A^4-B^4}{g^4\Gamma^4N^2_c}\Bigg(\Bigg[ \frac{N_c^6+2 N_c^4-19 N_c^2+8}{2(N_c^2-1)^2}-2\frac{N_c^6-3 N_c^4-26 N_c^2+16}{N_c^4-5 N_c^2+4}e^{-\frac{Q_{s\scaleto{1}{4pt}}^2r^2}{4}} \nonumber \\
&+(N_c^2-1)\!\left(1-e^{-\frac{Q_{s\scaleto{1}{4pt}}^2r^2}{4}}\left(1+\frac{Q_{s\scaleto{1}{4pt}}^2r^2}{4}\right)\right)\!\left(1-e^{-\frac{Q_{s\scaleto{2}{4pt}}^2r^2}{4}}\left(1+\frac{Q_{s\scaleto{2}{4pt}}^2r^2}{4}\right)\right)\nonumber \\
& +\frac{r^4}{4}Q_{s\scaleto{1}{4pt}}^2Q_{s\scaleto{2}{4pt}}^2-2r^2Q_{s\scaleto{1}{4pt}}^2\left(1-e^{-\frac{Q_{s\scaleto{2}{4pt}}^2r^2}{4}}\right)+ 2\frac{ \left(N_c^2-8\right) \left(N_c^2-1\right) \left(N_c^2+4\right)}{(N_c^2-4)^2}e^{-\frac{(Q_{s\scaleto{1}{4pt}}^2+Q_{s\scaleto{2}{4pt}}^2)r^2}{4}}\nonumber\\
& +\frac{(N_c-1)(N_c+3)N^3}{2(N_c+1)^2(N_c+2)^2}\left(\frac{N_c}{2}e^{-\frac{(N_c+1) r^2 Q_{s\scaleto{2}{4pt}}^2}{2 N_c}}+(N_c+2)-2(N_c+1)e^{-\frac{Q_{s\scaleto{2}{4pt}}^2r^2}{4}}\right)e^{-\frac{(N_c+1) r^2 Q_{s\scaleto{1}{4pt}}^2}{2 N_c}} \nonumber\\
& +\frac{(N_c+1)(N_c-3)N^3}{2(N_c-1)^2(N_c-2)^2}\left(\frac{N_c}{2}e^{-\frac{(N_c-1) r^2 Q_{s\scaleto{2}{4pt}}^2}{2 N_c}}+(N_c-2)-2(N_c-1)e^{-\frac{Q_{s\scaleto{2}{4pt}}^2r^2}{4}}\right)e^{-\frac{(N_c-1) r^2 Q_{s\scaleto{1}{4pt}}^2}{2 N_c}}\Bigg] \nonumber\\
& + \left[ 1\leftrightarrow2\right]\Bigg),\label{FullResult}
\ea
where the dependencies have been omitted for readability. Here we defined the saturation scale characterizing each nucleus as:
\begin{equation}\label{Saturation2}
\frac{r^2Q_s^2}{4}=g^2\frac{N_c}{2}\Gamma(r_{\pperp})\bar{\lambda}(b_{\pperp}).
\end{equation}
$A(r_{\pperp})$ and $B(r_{\pperp})$ were introduced in \eqref{param} as coefficients of the unpolarized and linearly polarized parts of the gluon distribution, respectively. Although we provided their general expressions in terms of $f(r_{\pperp})$ (\eqref{unpol} and \eqref{pol}), in order to analyze our result quantitatively we need to adopt an specific model. In the particular case of the strict MV model, where $f(r_{\pperp})$ is taken as a Dirac delta, we can compute these functions as:
\ba
A(r_{\pperp})_{\scaleto{\text{MV}}{0.12cm}}&=-\frac{1}{2}G(r_{\pperp})=\frac{1}{4\pi}K_{\scaleto{0}{4.2pt}}(mr)\label{ABMV1}\\
B(r_{\pperp})_{\scaleto{\text{MV}}{0.12cm}}&=\frac{1}{4\pi},\label{ABMV2}
\ea
where $K_{\scaleto{0}{4.2pt}}$ is a modified Bessel function and $m$ is the infrared cut-off mentioned in the previous section, now employed to regularize the divergent Green's function $G(r_{\pperp})$:
\ba
G(r_{\pperp})=\!-\!\int \frac{d^2q_{\pperp}}{(2\pi)^2}\frac{e^{iq_{\pperp}\!\cdot r_{\pperp}}}{q^2+m^2}=-\frac{1}{2\pi}K_{\scaleto{0}{4.2pt}}(m\,r).
\ea
Note that although we could have used an unrelated regulator mass, for the sake of simplicity we choose it to be the same one introduced before. In our calculation we will consider only the leading behavior in the $m\!\rightarrow\!0$ limit, which is:
\ba
A(r_{\pperp})_{\scaleto{\text{MV}}{0.12cm}}\approx\frac{1}{8\pi}\ln\!\left(\frac{4}{m^2r^2}\right)\!.
\ea
$B_{\scaleto{\text{MV}}{0.12cm}}$, being a constant, yields a negligible correction to this logarithm. In the same limit, the leading behavior of $\Gamma(r_{\pperp})$ corresponds to the following expression:
\ba
\Gamma(r_{\pperp})_{\scaleto{\text{MV}}{0.12cm}}=\frac{1}{2\pi m^2}-\frac{r}{2\pi m}K_{\scaleto{1}{4.2pt}}(mr)\approx\frac{r^2}{8\pi}\ln\left(\frac{4}{m^2r^2}\right).
\ea
Except for $B_{\scaleto{\text{MV}}{0.12cm}}$, all these factors exhibit logarithmic divergences. However, as all logarithms stemming from $A$ and $\Gamma$ are cancelled in the prefactor of \eqref{FullResult}, the only divergences that we need to deal with are the ones included in the saturation scale $Q_s^2$ (\eqref{Saturation2}), which diverges in both infrared $m\!\rightarrow\!0$ and ultraviolet $r\!\rightarrow\!0$ limits. Different prescriptions with a varying level of sophistication are available in the literature to treat this issue. In order to give a general idea of the magnitude and analytical features of our solution, we will adopt the GBW model, which in practice consists simply in neglecting all logarithmic dependencies. In this framework, on \figref{Result1} we draw the ratio of \eqref{FullResult} to the square of the energy density average:
\ba
\langle \epsilon_{\scaleto{0}{3.8pt}}(x_{\pperp})\rangle_{\scaleto{\text{\tiny MV}}{0.12cm}}\!=\frac{C_{\scaleto{F}{0.4em}}}{g^2}Q^2_{s\scaleto{1}{4pt}}Q^2_{s\scaleto{2}{4pt}}\,,
\ea
(whose computation can be found in \cite{Albacete:2018bbv}) as a function of the dimensionless product $rQ_s$ for $Q_{s\scaleto{1}{4pt}}\!=\!Q_{s\scaleto{2}{4pt}}$. Note that we are also taking $h(b_{\pperp})\!=\!1$ (strict MV model).

Although \eqref{FullResult} is somewhat lengthy, the following simplifying limits provide us with remarkably more compact formulas. In the limit of small transverse separations $r\!\to\!0$ the two-point function tends to:
\begin{align}
\lim_{r\rightarrow0}\,\langle\dot{\nu}_{\scaleto{0}{4.2pt}}(x_{\pperp})\dot{\nu}_{\scaleto{0}{4.2pt}}(y_{\pperp})\rangle_{\scaleto{\text{\tiny MV}}{0.12cm}}\!=\frac{3(N^2_c-1)}{32g^4N_c^2}Q_{s\scaleto{1}{4pt}}^4Q_{s\scaleto{2}{4pt}}^4.
\end{align}
The ratio with the product of the energy density averages at each transverse position reads:
\begin{align}
\lim_{r\rightarrow 0}\left(\frac{\langle\dot{\nu}_{\scaleto{0}{4.2pt}}(x_{\pperp})\dot{\nu}_{\scaleto{0}{4.2pt}}(y_{\pperp})\rangle}{\langle \epsilon_{\scaleto{0}{3.8pt}}(x_{\pperp})\rangle\langle \epsilon_{\scaleto{0}{3.8pt}}(y_{\pperp})\rangle}\right)_{\!\!\scaleto{\text{\tiny MV}}{0.12cm}}\!\!&=\frac{3}{8(N^2_c-1)},
\end{align}
which displays the characteristic $1/(N_c^2-1)$ suppression factor of non-trivial color correlations.
In the opposite limit, $rQ_s\gg1$, we obtain:
\begin{align}
\lim_{r Q_s\gg1}\!\left(\frac{\langle\dot{\nu}_{\scaleto{0}{4.2pt}}(x_{\pperp})\dot{\nu}_{\scaleto{0}{4.2pt}}(y_{\pperp})\rangle}{\langle \epsilon_{\scaleto{0}{3.8pt}}(x_{\pperp})\rangle\langle \epsilon_{\scaleto{0}{3.8pt}}(y_{\pperp})\rangle}\right)_{\!\!\scaleto{\text{\tiny MV}}{0.12cm}}\!\!&=\frac{32}{(N^2_c-1)^2r^4Q_{s\scaleto{1}{4pt}}^2Q_{s\scaleto{2}{4pt}}^2}.\label{LongLimit}
\end{align}
The previous expressions, being more `user-friendly' than our complete result, greatly simplify the potential application of this work to further analytical or phenomenological calculations. Also, they allow for a straightforward comparison of our approach to other analytical frameworks available in the literature, which is the main subject of the next subsection.

\subsection{The Glasma Graph approximation} 
\label{sec:gg}
The correlators presented in this paper, along with the energy density two-point function, were previously calculated in \cite{Lappi:2017skr} under the so-called Glasma Graph approximation. In this approach it is assumed that the four-point correlation functions of the gluon fields factorize into products of two-point correlation functions such that:
\begin{align}
\langle \alpha^{i,a}(x_{\pperp}) \alpha^{k,c}(x_{\pperp}) \alpha^{i^{\prime}\!,a^{\prime}}(y_{\pperp})\alpha^{k^{\prime}\!,c^{\prime}}(y_{\pperp}) \rangle_{\scaleto{\text{GG}}{0.12cm}}=\,&\langle\alpha^{i,a}(x_{\pperp}) \alpha^{k,c}(x_{\pperp})\rangle\langle \alpha^{i^{\prime}\!,a^{\prime}}(y_{\pperp})\alpha^{k^{\prime}\!,c^{\prime}}(y_{\pperp}) \rangle\nonumber\\
+ &\langle \alpha^{i,a}(x_{\pperp})\alpha^{i^{\prime}\!,a^{\prime}}(y_{\pperp})\rangle\langle \alpha^{k,c}(x_{\pperp})\alpha^{k^{\prime}\!,c^{\prime}}(y_{\pperp})\rangle\nonumber\\
+ &\langle\alpha^{i,a}(x_{\pperp})\alpha^{k^{\prime}\!,c^{\prime}}(y_{\pperp})\rangle\langle\alpha^{k,c}(x_{\pperp})\alpha^{i^{\prime}\!,a^{\prime}}(y_{\pperp})\rangle.\label{GGallin}
\end{align}
This Wick theorem-like decomposition is equivalent to assuming that the gluon fields conserve the Gaussian character of the color source distributions. This is not generally correct, as the dynamical generation of the former by the latter (encoded in the Yang-Mills equations) is non-linear. However, as observed in a previous work \cite{Albacete:2018bbv}, this assumption yields a good approximation of the exact result in the limit of small transverse separations $r\!\to\!0$. In this limit an effective linearization of the fields' dynamics takes place, as the connected function \eqref{connectedf} vanishes and the disconnected contributions become equivalent to the two-point function of gluon fields. 
This results in a mapping of the Gaussian statistics followed by the color source distributions onto the gluon fields.
\begin{figure}
\centering
\subfigure{\label{fig:2a}\includegraphics[width=0.510\textwidth]{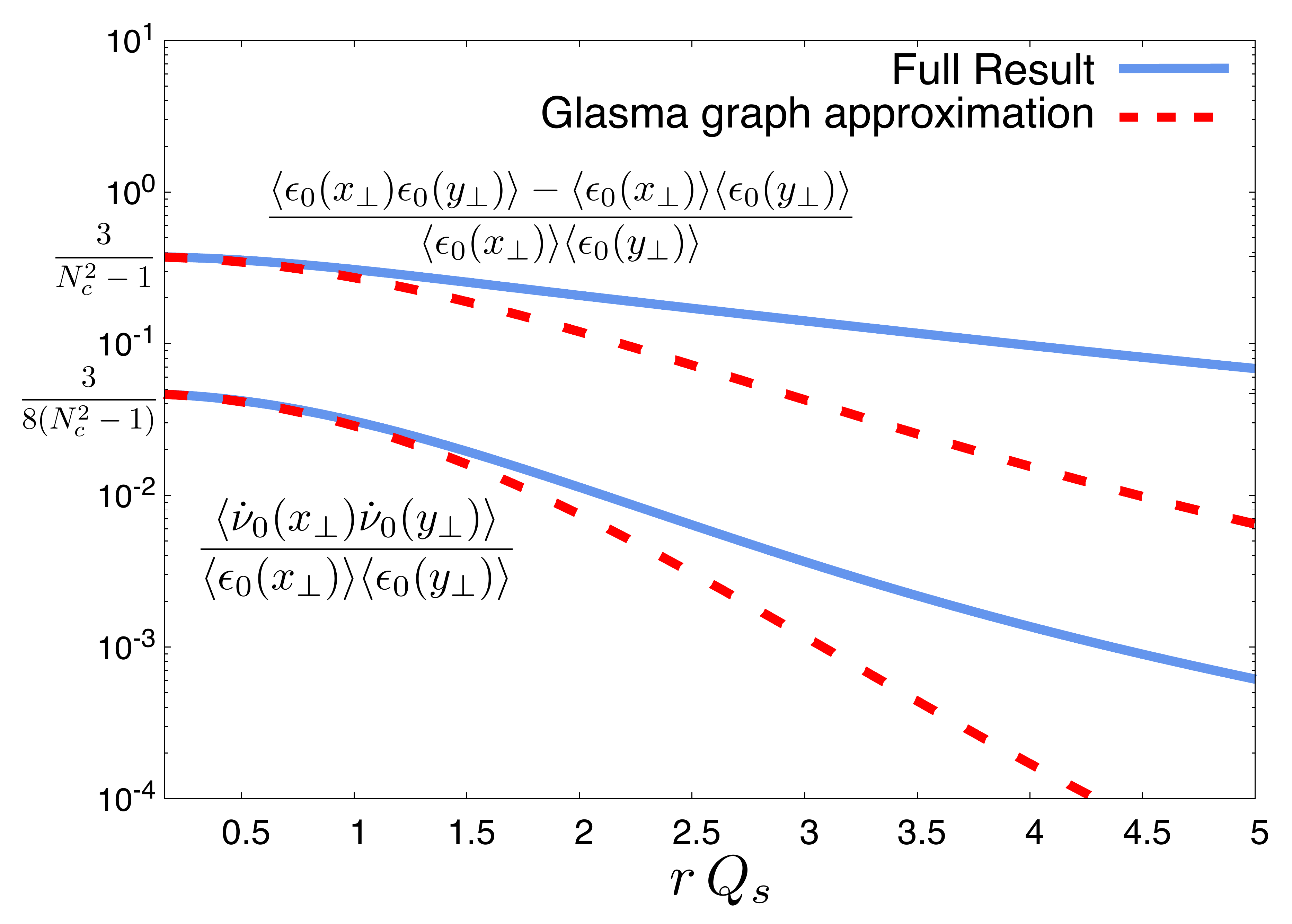}}
\subfigure{\label{fig:2a}\includegraphics[width=0.477\textwidth]{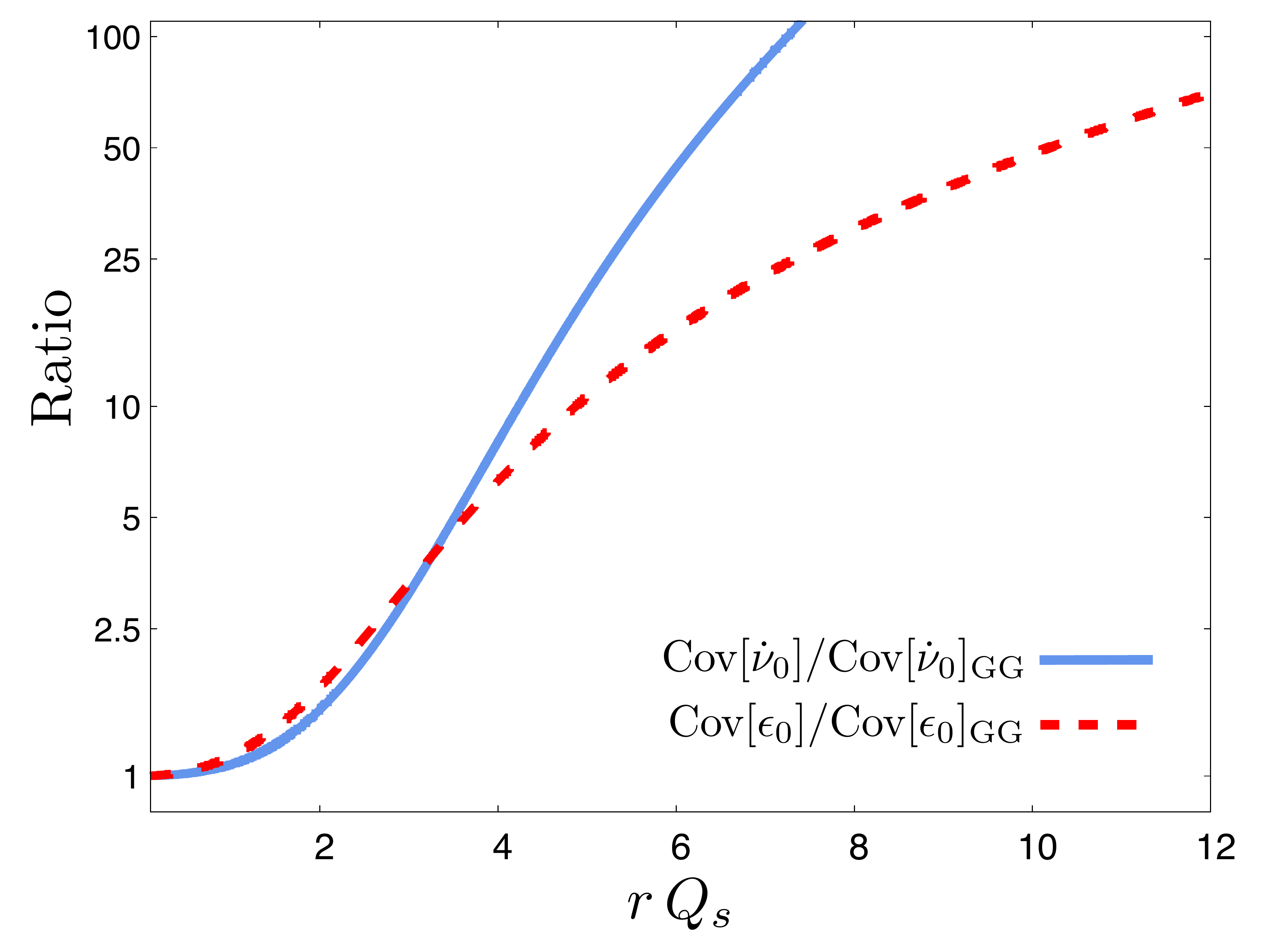}}
\caption{LEFT: Comparison of the covariance of the divergence of the Chern-Simons current (lower pair of curves) and the energy density (upper pair of curves) against $rQ_s$ for $Q_{s\scaleto{1}{4pt}}\!=\!Q_{s\scaleto{2}{4pt}}$ in the exact analytical approach (blue full curves) and the Glasma Graph approximation (red dashed curves). RIGHT: Ratio of exact analytical result to the Glasma Graph result for the covariance of the divergence of the Chern-Simons current (blue full curve) and the energy density (red dashed curve).}
\label{Result2}
\end{figure}

A comment is in order with respect to the designation of this approach. The original Glasma Graph approximation combines Gaussian statistics with the assumption that the valence quarks interact with the classical field by exchanging only two gluons, being applicable in the dilute limit \cite{Lappi:2015vta}. This results in a factorization of double parton distributions into all possible products of single parton distributions, which yields great simplification in the context of the calculation of di-hadron correlators \cite{Lappi:2017skr}. In the same spirit, the decomposition defined in \eqref{GGallin} proposes a similar approach to the calculation of $\langle\dot{\nu}(x_{\pperp})\dot{\nu}(y_{\pperp})\rangle$, which is thus expressed in terms of the building block defined for $\langle\dot{\nu}(x_{\pperp})\rangle$ (namely the two-point correlator of gluon fields). Ignoring conceptual differences, in this paper we will give the name `Glasma Graph approximation' to the approach based on said decomposition.

\begin{figure}[t]
\centering
\subfigure{\label{fig:2a}\includegraphics[width=0.496\textwidth]{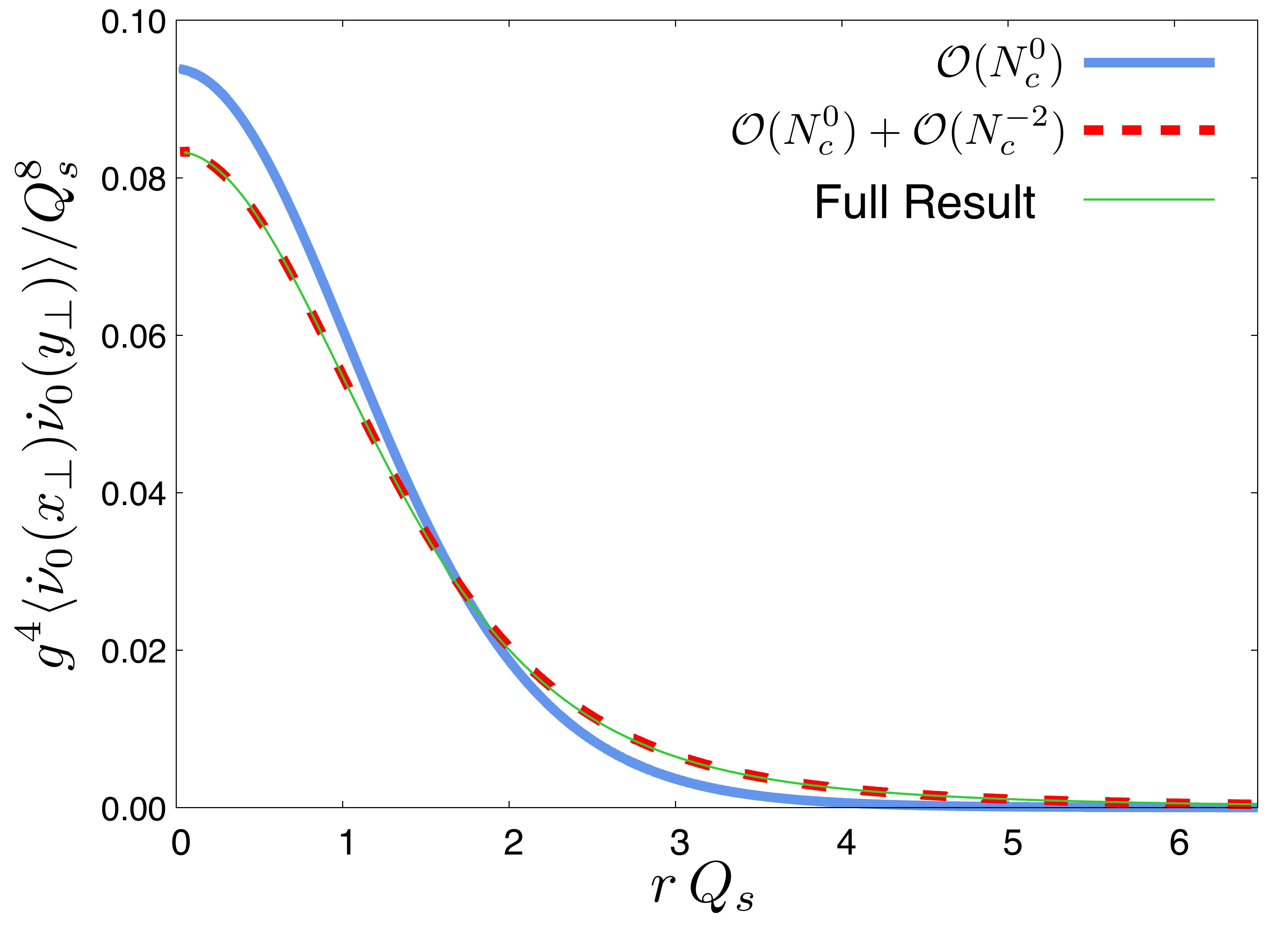}}
\subfigure{\label{fig:2a}\includegraphics[width=0.4955\textwidth]{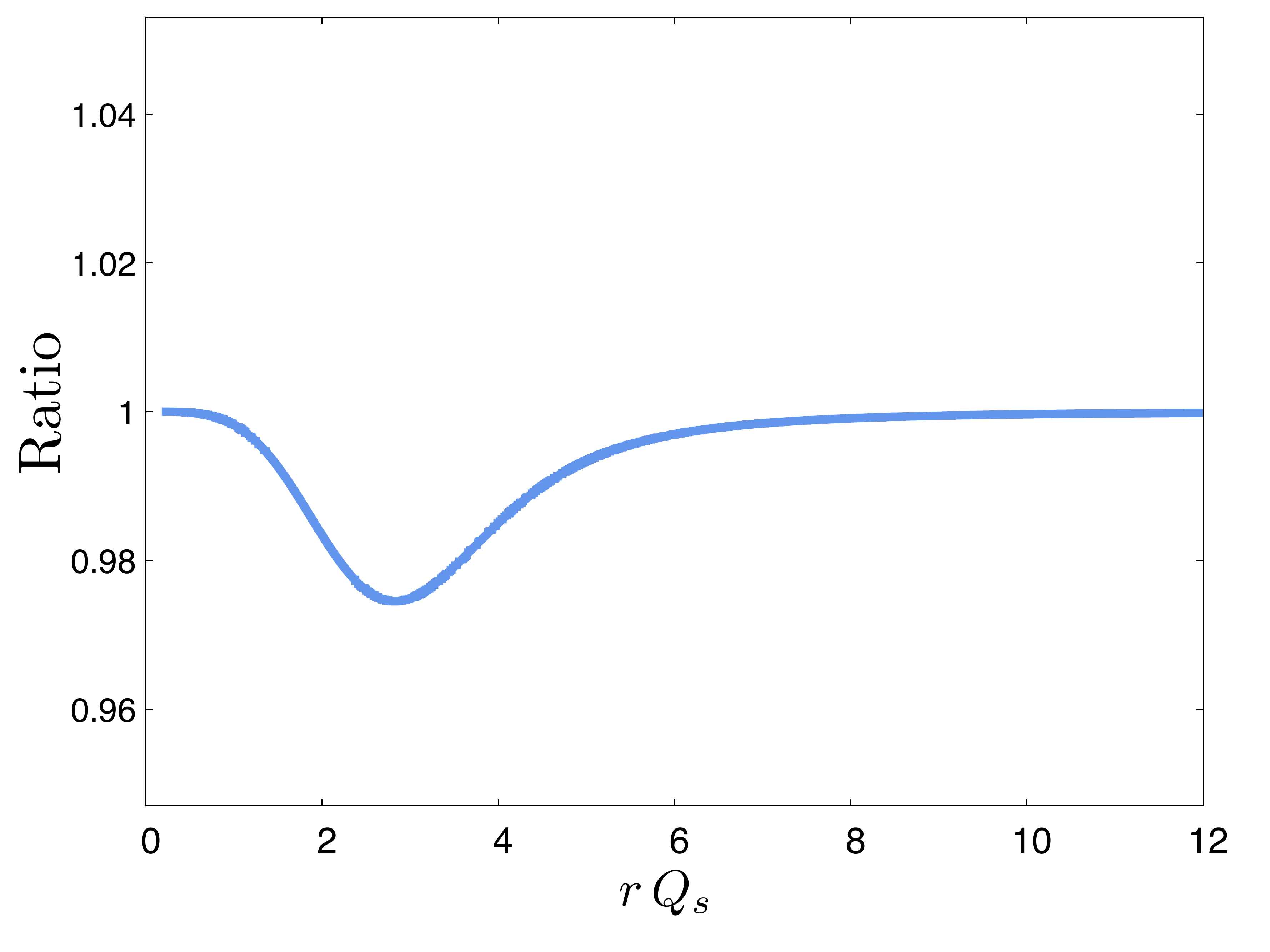}}
\caption{LEFT: Comparison of the first two orders of the $N_c$-expansion of the two-point function of the divergence of the Chern-Simons current against $r\,Q_s$ in the GBW model for $Q_{s\scaleto{1}{4pt}}\!=\!Q_{s\scaleto{2}{4pt}}$ and $N_c\!=\!3$. Blue full curve: $N_c^0$-order term. Red dashed curve: Sum of $N_c^0$- and $N_c^{-2}$-order terms. Thin green curve: full result. RIGHT: Ratio between the full result and the sum of the first two orders of the $N_c$-expansion.}
\label{Result3}
\end{figure}

In \figref{Result1} and \figref{Result2} we compare our result (in the strict MV model and with $Q_{s\scaleto{1}{4pt}}\!=\!Q_{s\scaleto{2}{4pt}}$) with the one computed according to the Glasma Graph approximation. As can be seen in \figref{Result1}, although both results agree exactly in the small transverse separation limit $r\!\rightarrow\!0$, in the rest of the spectrum (approximately for $r\!>\!1/Q_s$) our computation yields a significantly harder curve. As shown in \figref{Result2}, this is also the case for the two-point function of the energy density, computed in the exact analytical approach in \cite{Albacete:2018bbv}. Another major difference observed in said paper --and one of its main results-- consists in a comparatively slow vanishing behavior in the infrared limit, where the covariance of the energy density decreases following a $\sim\!1/r^2$ asymptotic curve, in stark contrast with the much steeper $\sim\!1/r^4$ decreasing behavior displayed by the Glasma Graph result. Remarkably, in the calculation presented here this difference becomes larger (as can be seen in the right panel of \figref{Result2}). In the present case, while our result shows a $\sim\!1/r^4$ decreasing behavior (see \eqref{LongLimit}), the Glasma Graph approximation yields a much steeper $\sim\!1/r^8$ tail:
\begin{equation}
\lim_{rQ_s\gg1}\left(\frac{\langle\dot{\nu}_{\scaleto{0}{3.8pt}}(x_{\pperp})\dot{\nu}_{\scaleto{0}{3.8pt}}(y_{\pperp})\rangle}{\langle\epsilon_{\scaleto{0}{3.8pt}}(x_{\pperp})\rangle\langle\epsilon_{\scaleto{0}{3.8pt}}(y_{\pperp})\rangle}\right)_{\!\!\scaleto{\text{GG}}{0.3em}}\!=\frac{96}{(N_c^2-1)r^8Q_{s\scaleto{1}{4pt}}^4Q_{s\scaleto{2}{4pt}}^4}.
\end{equation}
This discrepancy potentially implies much different numerical results and physical interpretations for any observable built from this quantity, in particular those based in fluctuations of the global amount of axial charge per unit rapidity (as they are proportional to the two-dimensional transverse integral of \eqref{LongLimit}).

\subsection{$N_c$-expansion} 
\label{sec:mv}
In order to complete the analysis of our final expression \eqref{FullResult}, in this subsection we will show its expansion in orders of $N_c$. The leading order term, of order $N_c^0$, reads:
\begin{align}
\left[\langle\dot{\nu}_{\scaleto{0}{3.8pt}}(x_{\pperp})\dot{\nu}_{\scaleto{0}{3.8pt}}(y_{\pperp})\rangle\right]_{N_c^0}=\left[\frac{1}{g^4\,r^8}e^{-\frac{r^2}{2} \left(Q_{s\scaleto{1}{4pt}}^2+Q_{s\scaleto{2}{4pt}}^2\right)}\!\left(8+16 e^{\frac{Q_{s\scaleto{1}{4pt}}^2 r^2}{2}}\!-32e^{\frac{Q_{s\scaleto{1}{4pt}}^2 r^2}{4}}\!+\,24 e^{\frac{r^2}{2} \left(Q_{s\scaleto{1}{4pt}}^2+Q_{s\scaleto{2}{4pt}}^2\right)}\right.\right.\nonumber\\
\left.\left.-8 e^{\frac{r^2}{4} \left(2 Q_{s\scaleto{1}{4pt}}^2+Q_{s\scaleto{2}{4pt}}^2\right)}\!\left( 8+Q_{s\scaleto{2}{4pt}}^2r^2\right)+e^{\frac{r^2}{4} \left(Q_{s\scaleto{1}{4pt}}^2+Q_{s\scaleto{2}{4pt}}^2\right)}\!\Big(Q_{s\scaleto{1}{4pt}}^2 Q_{s\scaleto{2}{4pt}}^2 r^4+4r^2(Q_{s\scaleto{1}{4pt}}^2+Q_{s\scaleto{2}{4pt}}^2)+48\!\Big)\right)\right]\!+\left[1\leftrightarrow2\right],\label{LargeN-NLO}
\end{align}
and the next term, of order $N_c^{-2}$, reads:
\begin{align}
\left[\langle\dot{\nu}_{\scaleto{0}{3.8pt}}(x_{\pperp})\dot{\nu}_{\scaleto{0}{3.8pt}}(y_{\pperp})\rangle\right]_{N_c^{-2}}=\left[\frac{1}{N_c^2g^4\,r^8}e^{-\frac{r^2}{2} \left(Q_{s\scaleto{1}{4pt}}^2+Q_{s\scaleto{2}{4pt}}^2\right)}\bigg(2\,Q_{s\scaleto{1}{4pt}}^2r^2(8+Q_{s\scaleto{1}{4pt}}^2r^2)e^{\frac{Q_{s\scaleto{2}{4pt}}^2 r^2}{2}}\right.\nonumber\\
+8\, e^{\frac{r^2}{4} \left(2Q_{s\scaleto{1}{4pt}}^2+Q_{s\scaleto{2}{4pt}}^2\right)} \Big( 4Q_{s\scaleto{1}{4pt}}^2 r^2+Q_{s\scaleto{2}{4pt}}^2 r^2-4\Big)+4e^{\frac{r^2}{2} \left(Q_{s\scaleto{1}{4pt}}^2+Q_{s\scaleto{2}{4pt}}^2\right)}\!\Big(Q_{s\scaleto{1}{4pt}}^2 Q_{s\scaleto{2}{4pt}}^2 r^4-4r^2(Q_{s\scaleto{1}{4pt}}^2+Q_{s\scaleto{2}{4pt}}^2)+4\Big)\nonumber\\
-4(Q_{s\scaleto{1}{4pt}}^4 r^4+12Q_{s\scaleto{1}{4pt}}^2 r^2+32)e^{\frac{Q_{s\scaleto{2}{4pt}}^2 r^2}{4}}-e^{\frac{r^2}{4} \left(Q_{s\scaleto{1}{4pt}}^2+Q_{s\scaleto{2}{4pt}}^2\right)}\!\Big(Q_{s\scaleto{1}{4pt}}^2 Q_{s\scaleto{2}{4pt}}^2 r^4+4r^2(Q_{s\scaleto{1}{4pt}}^2+Q_{s\scaleto{2}{4pt}}^2)-80\Big)\nonumber\\
+\left(Q_{s\scaleto{1}{4pt}}^2r^2+Q_{s\scaleto{2}{4pt}}^2r^2+8\right)^2\bigg)\!\bigg]\!+\left[1\leftrightarrow 2\right].\label{LargeN-NNLO}
\end{align}
As it is also the case for the covariance of the energy density $\epsilon_0$, the first two orders of the $N_c$-expansion of \eqref{FullResult} yield a neat approximation of the complete result (see \figref{Result3}), but not a significant improvement regarding the practicality of the formulas.

\section{Discussion and outlook} 
\label{sec:end}
In this paper we performed a first-principles analytical calculation of the two-point correlator of the divergence of the Chern-Simons current. This object characterizes a source of fluctuations of axial charge density in the Glasma state produced in the initial stage of an ultra-relativistic HIC. With this calculation we expand on the results presented in a previous work \cite{Albacete:2018bbv}, where we computed the covariance of the Glasma energy-momentum tensor. We performed both calculations following a classical approach based on the CGC effective theory, which we introduced by summarizing the computation of the gluon fields produced at $\tau\!=\!0^+$.  Our framework features an explicit impact parameter dependence in the two-point correlator of color source densities (first introduced in \cite{Chen:2015wia}), as well as a generalization of the transverse profile of the interaction. These modifications were also incorporated in \cite{Albacete:2018bbv} with the aim of expanding the potential phenomenological applications of our results. In the present work, however, we limited our analysis to the GBW prescription within the original MV model for simplicity.

With this setup we compare our result for the two-point correlator with the one obtained under the Glasma Graph approximation \cite{Lappi:2017skr}. As was also the case for the energy-momentum tensor \cite{Albacete:2018bbv}, the exact computation shows complete agreement with the Glasma Graph result in the ultraviolet $r\!\rightarrow\!0$ limit. However, a strong discrepancy emerges in the infrared $rQ_s\!\gg\!1$ limit: the exact two-point correlator of $\dot{\nu}$ (normalized to the product of the average energy densities at each transverse position) decays following a $\propto\!1/r^4$ power-law tail, whereas the Glasma Graph expression exhibits a much more rapidly decaying $\propto\!1/r^8$ behavior. Remarkably, the gap between both results is even larger than the one showed by the two-point correlator of the energy density, which in this limit disagrees with the Glasma Graph result by `only' a $1/r^2$ factor. This suggests that the non-linear dynamics followed by the gluon fields have an even greater effect over the long-range transverse fluctuations of axial charge density than they do over those of the deposited energy. On the other hand, the results show that for both calculations the fact that the Glasma field correlators do not obey Gaussian dynamics can be overlooked in the ultraviolet limit, or to a good approximation for correlation distances shorter than $1/Q_s$. This outcome agrees with the expected validity range of the Glasma Graph approximation \cite{Lappi:2015vta}.

One feature of our previous work \cite{Albacete:2018bbv} that is not reproduced by the results of the present paper is the logarithmic enhancement exhibited by the correlation length. The computation of the two-dimensional transverse integral of \eqref{LongLimit} is dominated by the lower bound $r\!\sim\!1/Q_s$, as opposed to the case of the corresponding energy density correlator, which is sensitive to the infrared cut-off $r\!\sim\!1/m$ via a logarithmic factor $\ln(Q_s/m)$. This result thus seems somewhat more consistent with the conjectured Glasma flux tube picture, which predicts the range of the transverse color screening of correlations to be of size $1/Q_s$ \cite{Iancu:2002aq}. Nevertheless, \eqref{LongLimit} still displays a remarkably slow fall-off that contrasts with the behavior one could naively expect from correlations between Gaussianly-distributed color charges.

The results of this work add further evidence on the importance of the non-linear dynamics relating color source densities and gauge field correlators beyond the validity range of the Glasma Graph approximation (thus supporting the conclusions reached in \cite{Albacete:2018bbv} in this regard). In addition, the expressions presented in this paper provide analytical insight on the early-time local fluctuations of axial charge density in the transverse plane. By following the practical steps first presented in \cite{Lappi:2017skr}, our formulas can be directly applied in phenomenological studies of anomalous transport phenomena. From \eqref{chiral} (rewritten using the covariant coordinate system, $(\tau,\eta,x_{\pperp})$), we obtain:
\ba
\frac{dN_5}{d^2x_{\pperp}d\eta }=\!\int\!d\tau\,\tau \partial_{\mu}j^{\mu}_5=\frac{g^2N_f}{2\pi^2}\!\int\!d\tau\,\tau \dot{\nu}(x).
\ea
Taking the first order of the $\tau$-expansion of $\dot{\nu}$ and integrating, we get to the differential distribution of axial charge density at early times:
\ba
\frac{dN_5}{d^2x_{\pperp}d\eta }\bigg\rvert_{\tau=0^+}\!\!=\frac{\tau^2}{2}\frac{g^2N_f}{2\pi^2}\dot{\nu}_{\scaleto{0}{3.8pt}}(x_{\pperp}),
\ea
From this expression one can straightforwardly relate the two-point function of the divergence of the Chern-Simons current computed here to the correlation function $\left\langle \frac{dN_5}{d^2x_{\pperp}d\eta }\frac{dN_5}{d^2y_{\pperp}d\eta }\right\rangle$. This object serves as the fundamental input for the Monte-Carlo modelization of initial conditions of axial charge density \cite{Lappi:2017skr}, required by those hydrodynamical simulations that aim at describing anomalous transport phenomena. The results based on the expressions presented here would of course be subject to higher order corrections in $\tau$. The computation of said terms, as well as the calculation of observables relevant to QGP phenomenology, are left for future work.



\acknowledgments

The author thanks Cyrille Marquet, Javier L.\ Albacete, Giuliano Giacalone and Tuomas Lappi for their useful feedback during the development of this work. Partial funding by a FP7-PEOPLE-2013-CIG Grant of the European Commission, reference QCDense/631558, the MINECO project FPA2016-78220 of the Spanish Government, and the `La Caixa' Banking Foundation is gratefully acknowledged by the author.

\bibliographystyle{JHEP-2modM}

\end{document}